# Exploration of reproducibility issues in scientometric research
# Part 1: Direct reproducibility

Ludo Waltman[*], Sybille Hinze[**], Andrea Scharnhorst[***], Jesper Wiborg Schneider[****], Theresa Velden[*****]

[*] *waltmanlr@cwts.leidenuniv.nl*
Centre for Science and Technology Studies (CWTS), Leiden University, Leiden (the Netherlands)

[**] *hinze@dzhw.eu*
Deutsches Zentrum für Hochschul- und Wissenschaftsforschung (DZHW), Berlin (Germany)

[***] *andrea.scharnhorst@dans.knaw.nl*
Data Archiving and Networked Services (DANS), Royal Netherlands Academy of Arts and Sciences, The Hague (the Netherlands)

[****] *jws@ps.au.dk*
Danish Centre for Studies in Research and Research Policy, Aarhus University, Aarhus (Denmark)

[*****] *velden@ztg.tu-berlin.de*
Zentrum für Technik und Gesellschaft (ZTG), Technische Universität Berlin, Berlin (Germany)

**Introduction**
The (lack of) reproducibility of published research results has recently come under close scrutiny in some fields of science (see e.g. Flier 2017 for a discussion of bio-sciences, and e.g. Open Science Collaboration 2015 and Pashler & Harris 2012 for an assessment of the situation in psychology). Aside from genuine error and fraud, theoretical investigations (e.g. Ioannidis 2005) and empirical investigations (e.g. John et al. 2012) point to the use of questionable research methods along with the overselling of results by overstating claims, and publication bias - the tendency to select positive results over negative results for publication - as important sources for the irreproducibility of published research.

In scientometrics we have not yet had an intensive debate about the reproducibility of research, although concerns about a lack of reproducibility have occasionally surfaced (see e.g. Glänzel & Schöpflin 1994 and Van den Besselaar et al. 2017), and the need to improve the reproducibility is used as an important argument for open citation data (see www.issi-society.org/open-citations-letter/). We initiated a first discussion about reproducibility in scientometrics with a workshop at ISSI 2017 in Wuhan.[1] One of the outcomes was the sense that scientific fields differ with regard to the type and pervasiveness of threats to the reproducibility of their published research, last but not least due to their differences in modes of knowledge production, such as confirmatory versus exploratory study designs, and differences in methods and empirical objects.

---

[1] Workshop report available online at www.issi-society.org/blog/posts/2017/november/reproducible-scientometrics-research-open-data-code-and-education-issi-2017/.

Therefore we suggest that an empirical investigation of the specific challenges to the reproducibility of research in the field of scientometrics would be beneficial to focus the debate and efforts to remedy shortcomings. As a first step, we decided to explore how we might assess 'in principle' reproducibility based on a critical review of the content of published papers. To this end we distinguish different categories of studies, and developed a taxonomy of threats to reproducibility that may be identified by a review of published papers. In part 1 (the paper you are currently reading), we focus on direct reproducibility - that is the exercise of a third party repeating a published study using the same method, data, procedures. In a companion paper, part 2 of this study (Velden et al. 2018), we focus on conceptual reproducibility - that is the exercise of a third party to test the robustness of knowledge claims of a study by reproducing the original claims using different data, methods, and procedures.

**Background**

The concept of reproducibility can refer to various approaches to and purposes of reproducing (some aspect of) an original study. What variety of reproducibility is seen as most pertinent, seems to depend on scientific domain. The diversity of perspectives has led to a thorough confusion of terminology around reproducibility, including antithetical definitions of the terms *replicability* and *reproducibility* (Goodman et al. 2016; Barba 2018). To cut through the thicket of terminological confusion, we use the term *reproducibility* as a generic umbrella term and focus on two distinct subtypes that we define as follows.

One way to think about differences between concepts of reproducibility is in terms of varying degrees of the similarity of conditions between the original study and a reproduction study, including the study design, methods, and data used (Chen 1994). We here distinguish two subtypes that are located at opposite ends of this spectrum and have distinct scientific functions: *direct* and *conceptual* reproducibility (in line with Fidler et al. 2017).

*Direct reproducibility* is located at the 'greatest similarity' end of the spectrum where the same data, tools and methods are used to reproduce and verify a study with the expectation of obtaining identical or very similar empirical results, unless some error is made either in the original study or in the reproduction study.

*Conceptual reproducibility* is located at the other end of the spectrum where a study is reproduced using different data, tools and methods with the aim of testing the robustness of the fundamental knowledge claims made by the original study. The concept is further discussed in the second part of our explorative analysis (Velden et al. 2018).

**Analytical approach**

To explore how one might identify reproducibility issues in publications of scientometric studies, we defined a categorization of types of scientometric studies and critically reviewed them with regard to potential threats to reproducibility. To ensure consistency across our reviews, we developed a taxonomy of potential threats to direct reproducibility, presented further below.

*Classification of studies*

First, we created a classification of scientometric studies into five categories in order to explore how threats to reproducibility may vary by type of study. We refined the empirical category in order to account for the large number and the large variety of empirical studies in scientometrics. Our classification is presented in Table 1. As often in classification, many studies do not fit neatly into one of our categories. Nevertheless, an assignment is possible by looking at the primary focus of a study.

Table 1: High-level classification of types of scientometric studies.

| Category no. | Name | Description |
| --- | --- | --- |
| 1 | Theoretical/Conceptual | Studies that are primarily theoretically/conceptually focused |
| 2 | Methods | Studies that are primarily methodologically focused. |
| 3 | Empirical (General) | Studies that are primarily empirically focused, aimed at answering substantive research questions in the study of science. |
| 4 | Empirical (Case) | Studies that are primarily empirically focused, taking a 'case study' approach, that is, focusing on analyzing particular research domains or particular countries, research institutions, or journals. These studies do not aim to develop more general insights that go beyond the particular case they analyze. |
| 5 | Empirical (Data Source) | Studies that are primarily empirically focused, aimed at getting a better understanding of the data sources available for scientometric research. |

*Taxonomy of threats to direct reproducibility*

A precondition for direct reproducibility is that the party attempting the reproduction has all the necessary means, information and resources (access to data, tools, infrastructures, relevant tacit knowledge). In a conflation of terminology, the ability to carry out a direct reproduction attempt is often not distinguished from the factual direct reproducibility of a study. In practice, a reproduction attempt may fail either because the preconditions for direct reproducibility are not met or because the original study is factually irreproducible. Given resource restrictions, we could not attempt direct reproduction, and therefore we restrict our review to the preconditions for direct reproducibility of the selected studies.

The taxonomy that we use captures threats to direct reproducibility by identifying issues that may practically undermine the ability of a third party to conduct a direct reproduction of a study. It distinguishes between critical dependencies (fundamental barriers that cannot be fixed by information provided in publication, such as access to original data or a certain tool used in the study) and issues resulting from incomplete information provided by the publication, either with regard to the procedures used, or with regard to the reporting of the results (see Figure 1).

Figure 1: Taxonomy to identify potential threats to direct reproducibility of a published scientometric study.

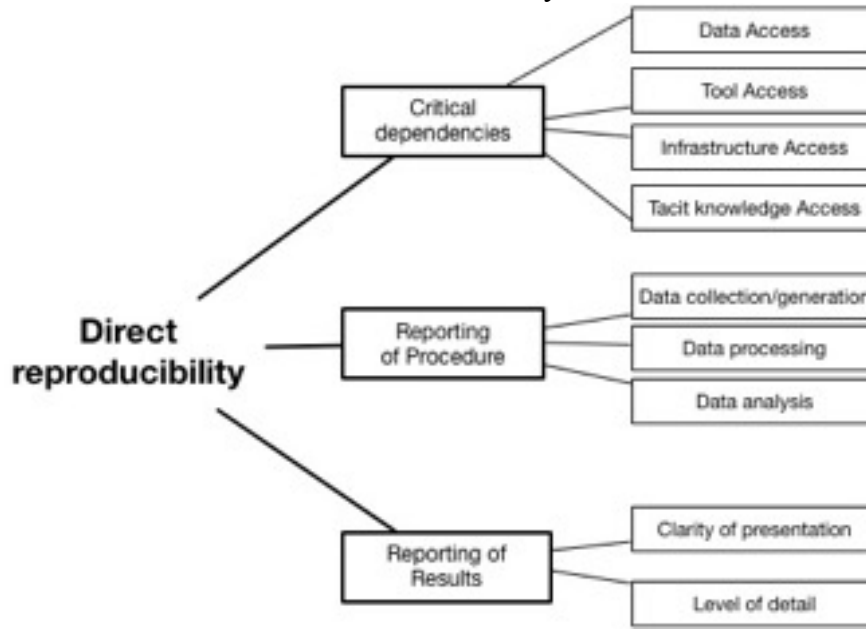

*Data and method*
For each of the five study type categories, we selected one paper that was published within the last two years. Two papers were published in *Scientometrics,* two in *Journal of Informetrics* and one paper was made available as a preprint in the arXiv. With the paper selection we aimed at selecting papers that serve as a good example of one the above five categories. Papers were selected and agreed upon unanimously by all authors of this paper. Each paper was then reviewed by at least two of the authors of this paper, one paper was reviewed by three. Each of the reviewers was asked to assess the papers regarding the elements identified by the taxonomy, for hurdles towards directly reproducing the research undertaken.

In this paper, we do not reveal the identity of the five papers, but we do provide an overview of key features of the papers in Table 2. Our focus is on providing general insights into the reproducibility of scientometric research, not about the extent to which specific papers are reproducible. Readers who want to know more about the papers that were reviewed are invited to contact us.

Table 2: Properties of the five papers selected for review in this explorative study.

| Paper no. | Study type | Topic area | Methods | Data | Tools |
|---|---|---|---|---|---|
| 1 | Theoretical/ conceptual | Citation theory | Theoretical reasoning, simulation | Synthetic | Self-developed simulation software |
| 2 | Method development | Topic extraction | Network clustering | Bibliometric, proprietary, large-scale ($10^7$) | Open source software |
| 3 | Empirical (Substantive) | Innovation studies | Statistical regression analysis, network analysis | Patent data, proprietary | Standard, proprietary statistical package, network analysis tool (proprietary, free trial) |
| 4 | Empirical (Case) | Specialty study at national level | Network analysis and visualization | bibliometric, proprietary, small-scale ($10^3$) | Freely accessible online tool |
| 5 | Empirical (Data source) | Evaluation of sources for citation analysis | Recall and precision measurements, correlation coefficients | Bibliometric, proprietary and freely accessible large-scale ($10^5$-$10^6$) | Freely accessible online tool for query generation |

**Results**

We organize our report of observations regarding direct reproducibility issues in four parts: Data, software tools, methods, and results.

*Data*

Empirical data was used in four of the five papers that we reviewed. In all four cases, the data was of a bibliometric nature. Scientometric studies may also use other types of data (e.g., data on peer review outcomes, data on research funding, or survey data), but no such studies were included in our analysis.

Basically, there seem to be two main problems with bibliometric data sources:
1. Some bibliometric data sources (e.g., Web of Science, Scopus, Derwent) are not freely accessible. Especially large-scale data access can be expensive, making it infeasible for many researchers to reproduce studies that rely on large-scale data access. Small-scale data access (e.g., through the web interfaces of Web of Science or Scopus, based on subscriptions) will often be less problematic and can be sufficient for scientometric case studies (category 4), but it is often insufficient for scientometric studies that aim to draw conclusions that are of general nature and that go beyond one specific case (category 3).
2. All bibliometric data sources seem to lack a systematic approach to version control. Papers sometimes indicate the date at which data was collected from a data source. This may be helpful to approximately reproduce the data collection, but it is not sufficient for exactly reproducing it. To exactly reproduce the data collection, data sources need to adopt a systematic approach to version control or authors need permission to share the primary data on which their study is based.

*Software tools*

We suggest that from the perspective of direct reproducibility it is useful to distinguish four levels of accessibility of software tools. These levels are listed in Table 3 in increasing order of the degree to which they support direct reproducibility of scientometric research. We found that the software tools used in the five papers reviewed cover all four levels of accessibility.

Table 3: Levels of accessibility to support direct reproducibility

| Access level | Description | Example | Implication |
|---|---|---|---|
| 0 | Custom software developed by the authors of a paper not made available to others | | Requires re-implementation |
| 1 | Commercial software | SPSS | Accessible only to those that can afford to use these tools |
| 2 | Freely available software, not open source | CiteSpace | Accessible to all; one has to rely on documentation for algorithmic details |
| 3 | Freely available software, open source | Gephi | Accessible to all; allows scrutiny of code for correctness and algorithmic details |

Another relevant issue is the distinction between short-term and long-term availability. We found that various software tools used in scientometric research are made available on personal websites, which does not seem to guarantee their long-term availability.

Finally, we note that some algorithms (e.g., clustering algorithms) implemented in software tools make use of computer generated pseudo random numbers. To achieve full reproducibility of the results, one needs to work with exactly the same random numbers. This means that the same random number generator with the same initial seed needs to be used. Software tools that do not support this will yield results that can be reproduced only in a statistical sense, and not in an exact sense.

*Methods*

In the case of all five papers that we reviewed, at least some of the reviewers expressed concerns about the lack of sufficient methodological details to enable full direct reproducibility.

Furthermore, although scientometric research relies mainly on quantitative methods, there sometimes also is a qualitative element in the methods, in particular when a quantitative scientometric method is evaluated qualitatively based on expert judgment. Full direct reproducibility of results obtained using qualitative methods may not be possible. For instance, different experts may have different opinions and even the same expert may not have the same opinion at two different points in time. Nevertheless, when qualitative methods are used, one may at least aim to make sure that the methods themselves are reproducible, even though this does not guarantee that the results will be fully reproducible as well.

*Results*

Two issues were identified related to the way in which the results of a study are reported.

First, results can be made available at different levels of aggregation. Papers tend to focus on reporting results at an aggregate level (e.g., distributions or summary statistics). This means that even if aggregate results have been successfully reproduced, it is not clear whether results at disaggregated levels have been reproduced as well. When it is considered desirable to reproduce the results of a study even at the most detailed level, results need to be available at this level.

Second, when detailed results are made available online in order to facilitate reproducibility, there is the issue of ensuring long-term availability of the results. This is similar to the issue of the long-term availability for software tools that was discussed above.

**Discussion**

This explorative study generated a number of open questions, offered for further consideration below.

A key issue relates to the trade-off between efforts invested in and potential benefits expected from improved direct reproducibility. How do we approach this cost-benefit trade-off? Does this trade-off vary by study type - e.g. do publications that produce (potentially) fundamental contributions to theory or method development deserve a higher level of effort to ensure direct reproducibility than publications of case studies with a limited scope and future applicability?

Another important question relates to the exact purpose of enhancing the direct reproducibility of scientometric research: For instance, is the purpose to screen for error or potential fraud, or is it to allow a third party to build confidence in the reported results by independently reproducing the study? Depending on the exact purpose, efforts made to enhance direct reproducibility may need to be focused in different ways.

Our explorative review suggests that certain issues related to direct reproducibility can be addressed by authors merely improving the reporting of their studies. However, complex procedures that require a lot of detail for full documentation and tacit components difficult to convey in writing constitute an important challenge.

Beyond improvements in reporting, more contentious is the question what to expect in terms of sharing material resources: the (oftentimes proprietary) primary data used, and the software and tools developed to conduct analyses. Here a number of concerns intersect:

1. What is really needed to enable the direct reproducibility of a study?
   a. When is the ability of inspect source code required, and under what conditions can software tools be accepted to reliably function as black boxes (e.g. a standard statistical analysis tool, a visualization tool etc.)?
   b. When is access to detailed result required? In case of the use of proprietary primary data, also the detailed result underlying the analysis often cannot be shared.
2. When are costs a third party would incur to reproduce a study, e.g. the cost of reimplementing an essential piece of software or infrastructure or of buying a large-scale proprietary data set, seen as prohibitive and what can be done about it?
3. How should the original team's effort (and perhaps also its potential sacrifice of 'competitive advantage') involved in enabling direct reproducibility be balanced against the investment needed to be made by another team to directly reproduce the original study?
4. What should be our expectations regarding the durability of access to tools and data that enable direct reproducibility? Is ad-hoc archiving and provision of access through personal websites sufficient, or should we develop strong recommendations towards the use of certified archiving services?

*Limitations*

This explorative study is only a first start to empirically ground an assessment of threats to reproducibility in scientometric research and to identify critical questions to be resolved in order to operationalize reproducibility for our field. The small, hand-selected sample of publications we reviewed is not representative for the entire body of research published in scientometrics, e.g. in terms study designs, methods, and data used. We aimed to capture some of the variety of studies we encounter in scientometrics by our categorization of fundamental study types. However, due to the smallness of the sample we could not capture the variation in methods and in study quality within each category of study types. Hence we cannot make conclusive statements on the extent to which reproducibility shortcomings may vary by study type. We only see initial pointers towards studies being under a threat of potentially not being reproducible, e.g. due to dependence on proprietary data as well as weaknesses in documenting and justifying methodological choices and decisions.

**Conclusion**

For the upcoming STI2018 conference, we suggest to discuss some of the questions raised in this paper. One of the key questions is the trade-off between benefits and costs of improving the direct reproducibility of published research. Can we identify specific areas or instances where lack of direct reproducibility has undermined scientific progress in scientometrics? How could this have been prevented? And what would have been the benefits and costs of preventing this?

Scientometric journals along with the peer review process as gatekeepers of what gets formally published in our field, are in a key position to set standards for best practices. Hence in terms of practical outcomes, we might consider taking steps towards developing guidelines for journal editors, reviewers and authors on good practices to ensure and promote direct reproducibility of published research.